\documentclass[pdflatex,sn-mathphys]{sn-jnl}

\usepackage{siunitx}

\jyear{2025}%

\theoremstyle{thmstyleone}%
\theoremstyle{thmstyletwo}%

\theoremstyle{thmstylethree}%

\begin{document}

\title[Virp]{Virp: neural network-accelerated prediction of physical properties in site-disordered materials}


\author*[1]{\fnm{Andy Paul} \sur{Chen}}\email{andypaul.chen@ntu.edu.sg}

\author[2]{\fnm{Martin Hoffmann} \sur{Petersen}}\email{mahpe@dtu.dk}

\author*[1,3]{\fnm{Kedar} \sur{Hippalgaonkar}}\email{kedar@ntu.edu.sg}

\affil*[1]{\orgdiv{School of Materials Science and Engineering}, \orgname{Nanyang Technological University}, \orgaddress{50
Nanyang Avenue, Singapore 639798, Republic of Singapore}}

\affil[2]{\orgdiv{Department of Energy Storage and Conversion}, \orgname{Technical University of Denmark}, \orgaddress{Anker Engelunds Vej 301, DK-2800 Kongens Lyngby, Denmark}}

\affil[3]{\orgdiv{Institute of Materials Research and Engineering}, \orgname{A*STAR - Agency for Science, Technology and Research}, \orgaddress{2 Fusionopolis Way, Singapore, 138634, Republic of Singapore}}


\abstract{Among metallic alloys, ceramics, and even common compounds such as water ice, it is usual to find materials in which crystalline order is expressed as a probability. In such cases, one or more sites within a crystal can be occupied by multiple elements or vacancies according to a set of probabilities. These crystal structures remain inaccessible to common first principles materials simulation methodologies, which assume perfect crystal order. Workaround strategies to this limitation include quasi-random structures and cluster expansion. These methods are system-specific and computationally expensive as they rely on large-scale Monte Carlo simulations of enlarged unit cells. To address these limitations, we propose a pipeline that combines a permutation-based virtual cell generation algorithm, sampling regime, and thermodynamic postprocessing which greatly improves the feasibility of computation analyses for site-disordered materials. We demonstrate that, as long as the supercell definition is sufficiently large, the massive configurational space can be adequately sampled with 400 virtual cells given an acceptable error margin of 5\%. In our tests, this translates to a dispersion of 0.03 g/cm$^3$ or below in predicted density figures.}




\maketitle

\section*{Introduction}\label{sec:intro}

Site-disordered materials are defined by crystal structures in which at least one crystallographic site is partially occupied by atoms of different elements. This category is diverse and encompasses materials from metal alloys such as CoFe\cite{Srivastava2018}, ordered vacancy compounds such as CuIn$_3$Se$_5$\cite{hanada_crystal_1997}, and correlated disorder materials such as water ice\cite{keen_crystallography_2015}. Site-disordered materials are distinguished from other modes of disorder such as the kind found in amorphous materials, grain boundaries, line defects, and plane defects. Point defects are a particular expression of site disorder. Stoichiometric tuning and doping in materials synthesis, by substituting one element for another at specific crystallographic sites, makes many synthetic compounds site-disordered\cite{prasanna_band_2017}. As such, site disorder plays an important role in enhancing functional properties in materials. Naturally occurring minerals are predominantly determined to exhibit site disorder. As such, site-disordered materials are ubiquitous in materials science.

\begin{table}[h]
\begin{center}
\caption{Prevalence of site-disordered materials in experimental databases}\label{tbl:dbstats}%
\begin{tabular}{@{}llll@{}}
\toprule
Database & Ordered & \textbf{Disordered} & Error \\
\midrule
AMCSD\cite{downs_american_2003} & 10655 (50.6\%) & \textbf{9153 (43.5\%)} & 1246 (5.9\%) \\
ICSD\cite{Bergerhoff1983} & 122517 (52.2\%) & \textbf{106970 (45.6\%)} & 4966 (2.2\%) \\
COD\cite{grazulis_crystallography_2009,grazulis_crystallography_2012} & 347025 (67.2\%) & \textbf{162321 (31.5\%)} & 6740 (1.3\%) \\
\botrule
\end{tabular}
\end{center}
\end{table}

First principles simulation methods, especially density functional theory (DFT), have been used in the past few decades to explore the link between crystal structure and material properties. Large DFT computational databases have been compiled, including Materials Project (MP)\cite{jain_commentary_2013} and the Open Quantum Materials Database (OQMD)\cite{Kirklin2015}. Despite their abundance in real life, site-disordered materials are conspicuously absent from these databases. This originates from the inability of DFT software to treat atomic sites with site disorder.

Numerous strategies exist to bridge this gap between theory and experiment. Cluster expansion and special quasirandom structures (SQS)\cite{gehringer_models_2023} are used to simulate random distributions of elements at the disordered sites in a quasirandom or virtual cell. The coherent potential approximation (CPA) method \cite{Vitos} is used to simulate an effective medium potential created by the mixture of elements in a disordered system. The applicability of these methods is limited to simple disordered materials, especially metal alloys, and SQS tends to be computationally expensive. Moreover, these methods are system-specific, making them inefficient for exploring several site-disordered crystals. In 2017, the software \texttt{Supercell}\cite{okhotnikov_supercell_2016} was the first to efficiently generate a large number of virtual supercells for materials that do not exhibit correlated disorder. More recently, in 2023, \texttt{aflow++}\cite{oses_aflow_2023} was developed, and it uses a batch of virtual supercells and DFT to predict the physical properties of a site-disordered material using Boltzmann averaging.

In these previous applications, the presence of site disorder turns one simple computational routine into many heavy routines, depending on the supercell size and the batch size. At this juncture, we are far from solving the problem of computational expense. However, current neural network-derived methods, such as machine-learned interatomic potentials (MLIPs), can greatly reduce computational times relative to DFT-based methods. 

\begin{figure}[htp]
\centering
\includegraphics[width=\textwidth]{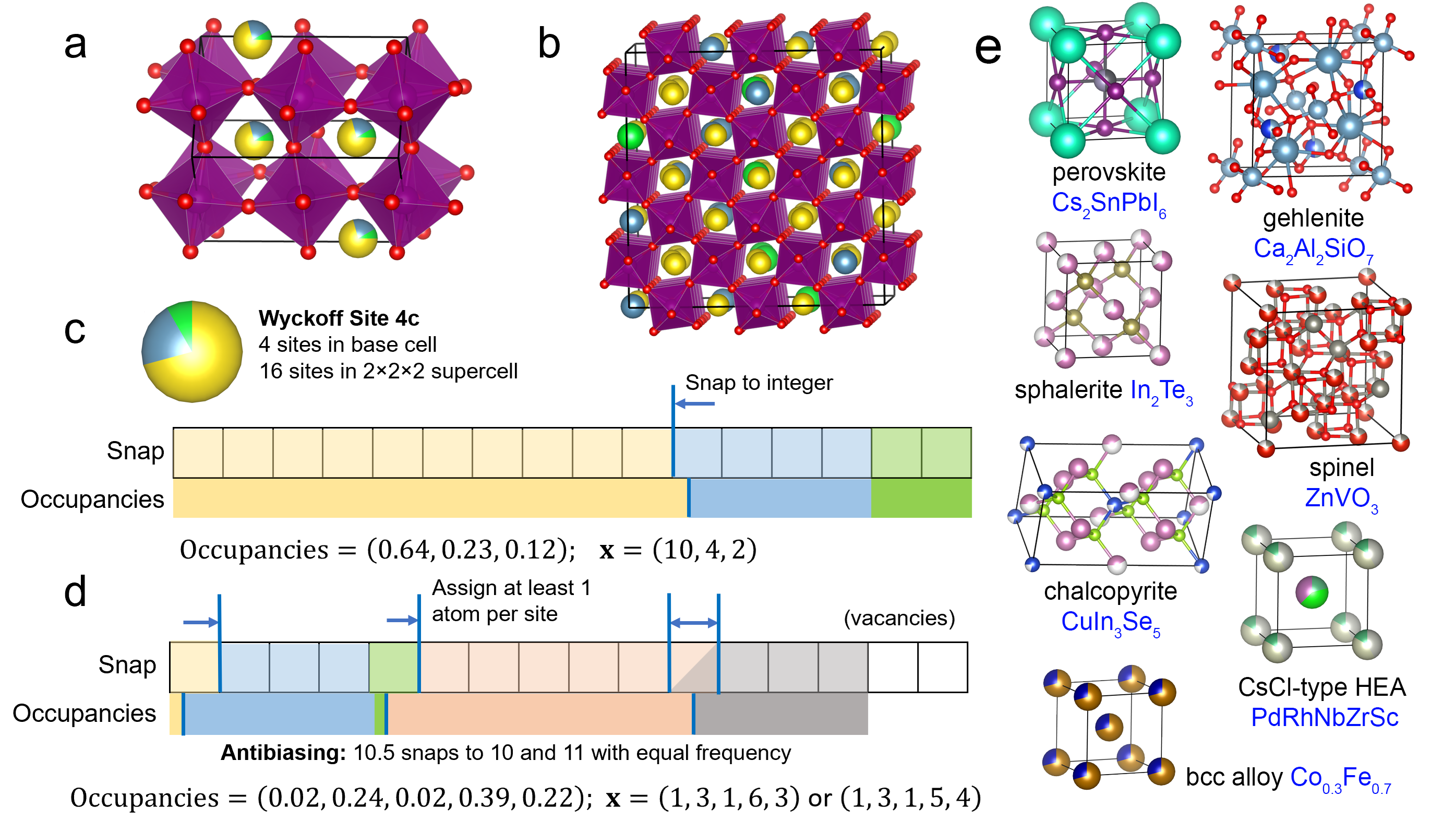}
\caption{The \texttt{virp} package workflow: from a site-disordered source unit cell (a) we generate a supercell (b). (c) The disordered site $4c$ is discretized by rounding into a snap array according to site occupancies and the site multiplicity in the supercell. (d) In another hypothetical disordered site, rounding exceptions require at least one atom of the element to be represented, and antibiasing mechanism averages stoichiometries in the case of exactly half-filled sites. The occupancy and snap vectors in these examples are presented. (e) Some examples of site-disordered materials, used in our trial demonstrations. Structure types in this paper (in black) and their corresponding chemical formulae (in blue) are specified.}\label{fig:virpintro}
\end{figure}

These improvements and best practices in current software applications are incorporated into the \texttt{virp} program. From site-disordered source cells, a lightweight and permissive procedure assigns atoms to disordered sites in a supercell, and creates DFT-treatable virtual cells closely approximating the stoichiometry of the source cell (Figure \ref{fig:virpintro}a-d). In contrast to the site occupation method employed in \texttt{pymatgen}\cite{ong_python_2013}, the virtual cell is not required to follow the exact stoichiometry of the source cell. This makes it useful for modeling large sets of site-disordered materials, including experimental series in which compositions can be adjusted along a continuous axis.

We incorporate MLIPs and a band gap model into our program to build large libraries of structurally optimized virtual cells and their predicted properties in a relatively short span of time. In addition, we consider and evaluate some computational load-saving strategies, including sampling the configurational space and identifying equivalent virtual cells. 

A selection of site-disordered crystal unit cells is used to demonstrate the database building operations of \texttt{virp}. We will refer to each sample from now on by their structure types, as detailed in Figure \ref{fig:virpintro}e. The trial set covers a diverse range of compounds and structure types, including perovskite Cs$_2$SnPbI$_6$\cite{mishra_synthesis_2025}, gehlenite Ca$_2$Al$_2$SiO$_7$\cite{okhotnikov_supercell_2016}, spinel ZnVO$_3$\cite{dai_data-driven_2025}, sphalerite In$_2$Te$_3$\cite{verkelis_beta-modification_1972}, chalcopyrite CuIn$_3$Se$_5$\cite{hanada_crystal_1997}, bcc alloy Co$_{0.3}$Fe$_{0.7}$\cite{hocine_role_2017}, and a CsCl-type high entropy alloy (HEA) PdRhNbZrSc\cite{stolze_sczrnbrhpd_2018}. Points of comparison can be made between highly symmetric cells (perovskite, bcc alloy, CsCl-type HEA) and asymmetric cells (gehlenite, chalcopyrite). The site occupancies range from a simple half-half split (e.g. perovskite) to more heterogeneous site allocations, which can also include vacancies (e.g. spinel). Crystal structures in which multiple sites exhibit disorder (chalcopyrite, CsCl-type HEA) are also considered in the trial set. In the following subsections, we demonstrate that virtual cells can be generated with ease, using \texttt{virp}.

\section*{Results}\label{sec3}

\subsection*{Determining the sample size}

\begin{figure}[h]
\centering
\includegraphics[width=\textwidth]{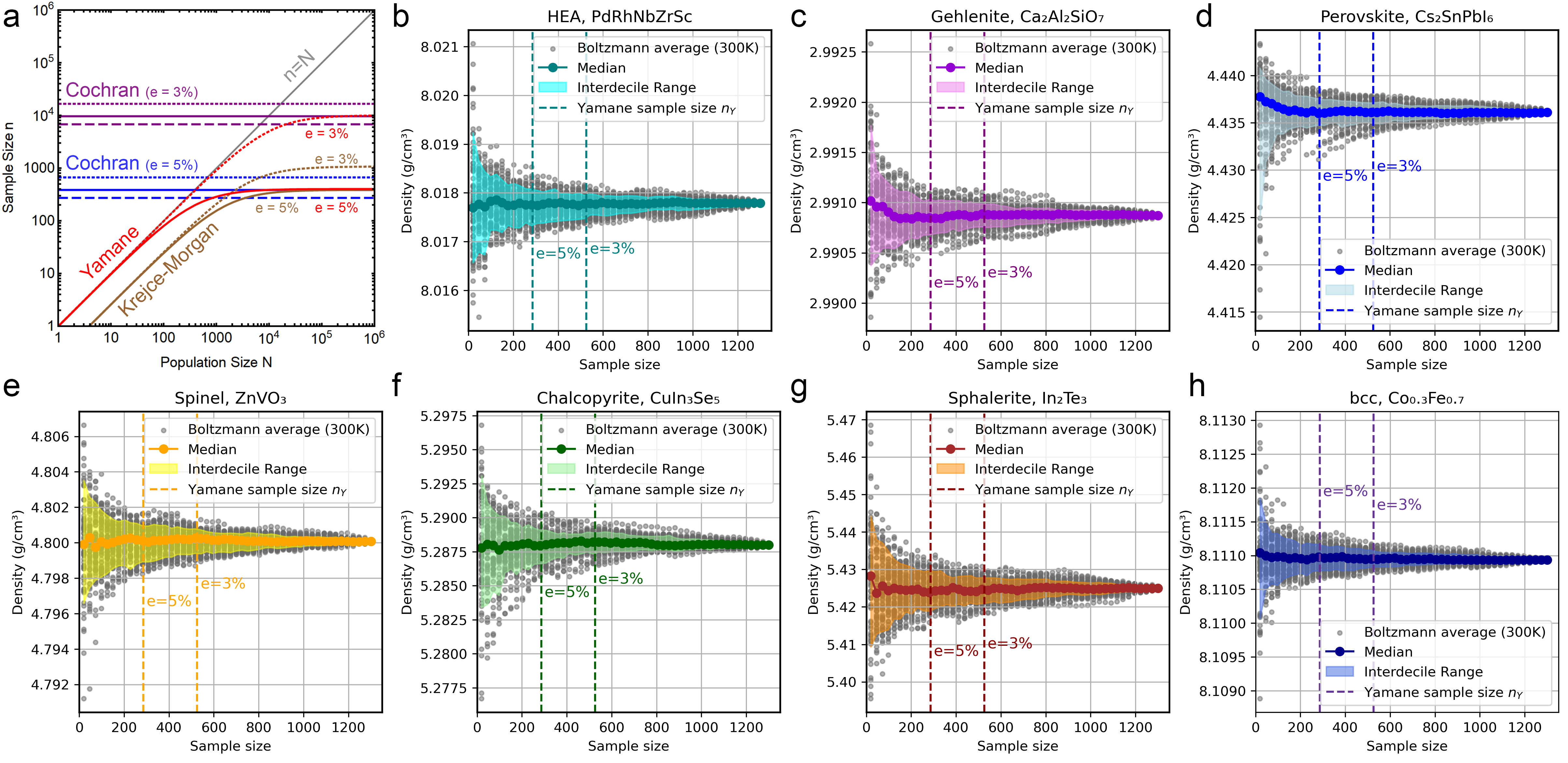}
\caption{(a) Boltzmann-averaged density of Co$_{0.3}$Fe$_{0.7}$ (bcc alloy) based on sampling a population of 1300 virtual cells; (b) Yamane, Cochran, and Krejce-Morgan sample sizes ($p=0.5$) against population size. For comparison, the measured density of Co$_{0.3}$Fe$_{0.7}$ is 8.017 g/cm$^3$\cite{hocine_role_2017}.}
\label{fig:yamane}
\end{figure}

The Yamane sampling regime is recommended for cases where the target quantity is continuous\cite{yamane_statistics_1967}. For a given population size $N$ and error margin $e$, the Yamane sample size $n_Y$ can be determined as:
\begin{equation}
    n_Y = \frac{N}{1+Ne^2}
\end{equation}
Despite its dependence on population size, $n_Y$ levels off for larger populations. Its upper bound can be expressed as  
\begin{equation}
    \lim_{N\rightarrow\infty} n_Y = \frac{1}{e^2} .
\end{equation}
Thus, for a chosen error margin of under 5\%, a sample size of 400 is sufficient to sample the configuration space for Boltzmann-averaged quantities, even as population sizes become very large (Figure \ref{fig:yamane}a). Contrary to what Ohkotnikov et al.\cite{okhotnikov_supercell_2016} and Oses et al.\cite{oses_aflow_2023} may suggest, a complete sampling of the configuration space is not necessary as long as one is interested in estimating the general crystal properties.

To test the error margins in practice and to its effect on the Boltzmann-averaged quantities, we generated 1300 virtual cells for each crystal structure in the trial set. Then, we define 50 sample sizes in a range from 20 to 1300, and the Boltzmann-averaged density of the structurally optimized cells is calculated 70 times for each sample size. The results are presented in Figure \ref{fig:yamane}b-h. These figures suggest that the Boltzmann-averaged density predictions from Yamane sampling stabilize around $n_Y$ samples (in this case, $n_Y=286$).

\begin{table}[h]
\begin{center}
\caption{Dispersion (interdecile range, IDR) of Boltzmann-averaged densities (in g/cm$^3$) of $n_Y=286$ samples of 1300 structures. The measured error margins are presented.}\label{tbl:configs}%
\begin{tabular}{@{}llll@{}}
\toprule
& IDR ($N$) & IDR ($n_Y$) & Error \\
\midrule
Perovskite & $2.1 \times 10^{-2}$ & $2.8 \times 10^{-2}$ & 13.1\% \\
Gehlenite & $7.1 \times 10^{-3}$ & $3.6 \times 10^{-4}$ & 5.0\% \\
Sphalerite & $2.1 \times 10^{-1}$ & $1.1 \times 10^{-2}$ & 5.2\% \\
Spinel & $3.9 \times 10^{-2}$ & $2.0 \times 10^{-3}$ & 5.2\% \\
Chalcopyrite & $2.1 \times 10^{-2}$ & $2.5 \times 10^{-3}$ & 11.5\% \\
bcc alloy & $2.9 \times 10^{-1}$ & $3.9 \times 10^{-4}$ & 13.7\% \\
CsCl-type HEA & $1.4 \times 10^{-2}$ & $7.5 \times 10^{-4}$ & 5.5\% \\
\botrule
\end{tabular}
\end{center}
\end{table}

Interestingly, the dispersion of Boltzmann-averaged values tends to be around 5\% or around 11-13\%. The higher dispersion of the perovskite density can be explained by the bimodality of the density originating from the antibiasing mechanism during site filling. Multimodality as a result of the high occurrence of vacancies can result in a higher dispersion of the predicted density in chalcopyrite. As Boltzmann averaging privileges structures with lower formation energies, the lowest energy virtual cells have a higher contribution to the predicted density compared to when only simple averaging is employed.

\subsection*{Periodic boundary image artifacts}

In any extended system modeled with a crystallographic unit cell, the influence of the periodic boundary must not be taken lightly \cite{Hine2009}. In a high-throughput problem setting such as disordered materials analysis, a rule of thumb should be defined such that database building routines can be performed with minimal human intervention. We compare the effects of supercell size against that of sample size on the thermoelectric material AgSbTe$_2$ \cite{roychowdhury_enhanced_2021}, with a prototypical disordered rocksalt structure with Te in the anionic sites and a 50/50 distribution of Ag and Sb atoms occupying the cationic sites. This choice is based on its relatively simple cubic symmetry and cationic ordering, which gives an isotropic distribution of the periodic boundaries, relatively few atomic species, and fewer complications from site vacancies.

\begin{figure}[htp]
\centering
\includegraphics[width=0.6\textwidth]{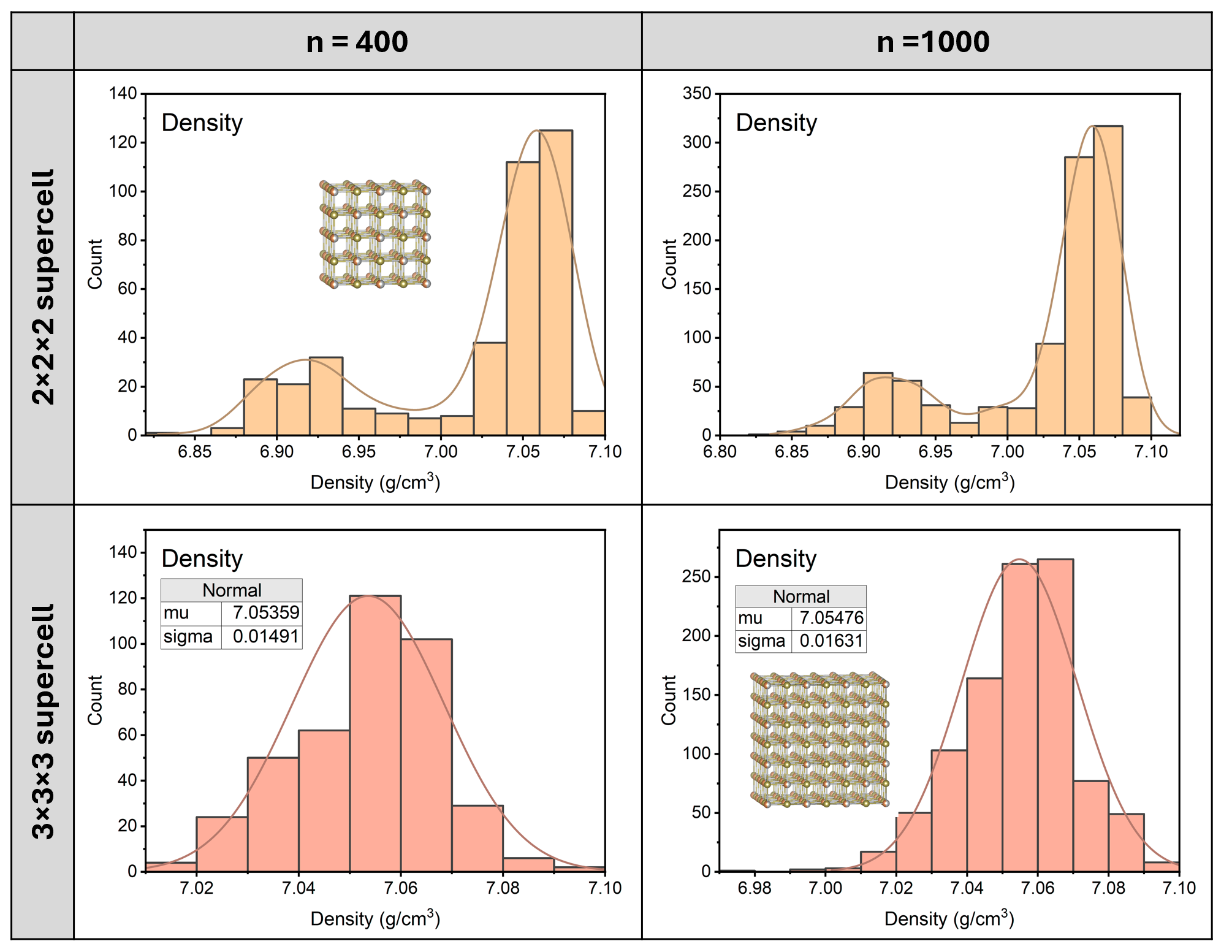}
\caption{Histograms of statistical distribution of crystallographic densities of supercells of AgSbTe$_2$. For reference, the corresponding (disordered) supercells are depicted on the right.}
\label{fig:pbe}
\end{figure}

Figure \ref{fig:pbe} examines the statistical distributions of the crystallographic density by comparing sets of $2 \times 2 \times 2$ and $3 \times 3 \times 3$ virtual cells with sample sizes of $n = 400$ and $n=1000$. For the $2 \times 2 \times 2$ virtual cells with $n=400$, a bimodal distribution can be seen in the density, with a lower peak around 6.90 g/cm$^3$ and a higher peak around 7.05 g/cm$^3$. A naive reading of the data would suggest that AgSbTe$_2$ could exist as a mixture of two crystallographic phases, each with their own set of lattice parameters and unit cell volumes. However, the same distribution for the larger $3 \times 3 \times 3$ virtual cells is no longer bimodal.

Moreover, increasing the sample size from 400 to 1000 supercells does not drastically change the character of the statistical distribution of the crystallographic densities for both supercells. This constitutes a strong argument that the choice of a sufficiently large supercell is more consequential than choosing a sample size much larger than that prescribed by the Yamane formula in the prediction of materials properties through Boltzmann averaging.

Since the periodic boundary effects are weaker in the larger ($3 \times 3 \times 3$) supercell, and all other factors are the same, we can conclude that the bimodality in $2 \times 2 \times 2$ supercells is purely an effect of interactions across the periodic boundary. 

From this example, one might adopt a general rule of thumb that a safe minimum distance between periodic boundary images when deciding the supercell size would be about 15$\si{\angstrom}$. This assumption is comparable to similar rules of thumb that govern the distance between low-dimensional models, such as 2D slab layers, commonly used in DFT\cite{Chen2020a}.

\subsection*{Symmetrical equivalence of virtual cells}

Accurate approximation of the disordered structure requires larger supercells. The probability that two generated virtual cells are symmetrically equivalent is small, since the symmetry of the original cell is broken during the supercell-generating and randomized site-filling operations. Even so, symmetrically-equivalent virtual cells may still occur if a pair of two different configurations happen, by chance, to produce cells which are congruent with each other through translation or rotation. These redundant cells can be identified by the CHGNet total energies of their relaxed structures without the need for symmetry resolution, which is computationally expensive. This is distinct from the approach of \texttt{Supercell}, which requires symmetry resolution and practically limits the size of the supercell one can choose.

\begin{table}[h]
\begin{center}
\caption{Size of configuration space (N) and redundancy in a set of 700 generated virtual structures from the trial set.}\label{tbl:configs}%
\begin{tabular}{@{}lllll@{}}
\toprule
& Supercell & N & Repeat & (\%) \\
\midrule
Perovskite & $3 \times 3 \times 3$ & $1.9 \times 10^{15}$ & 41 & 5.9\% \\
Gehlenite & $2 \times 2 \times 3$ & $6.4 \times 10^{27}$ & 20 & 2.9\% \\
Sphalerite & $3 \times 3 \times 3$ & $5.8 \times 10^{28}$ & 2 & 0.29\% \\
Spinel & $2 \times 2 \times 2$ & $2.4 \times 10^{46}$ & 6 & 0.86\% \\
Chalcopyrite & $3 \times 3 \times 2$ & $3.2 \times 10^{16}$ & 18 & 2.6\% \\
bcc alloy & $4 \times 4 \times 4$ & $5.0 \times 10^{32}$ & 34 & 4.9\% \\
CsCl-type HEA & $5 \times 5 \times 5$ & $1.7 \times 10^{110}$ & 13 & 1.9\% \\
\botrule
\end{tabular}
\end{center}
\end{table}

In our trial set materials, we found that the number of energetically degenerate virtual cells (i.e. the degeneracy) only goes up to almost $\sim$6\% (Table \ref{tbl:configs}) of the sample set. Note that certain high-symmetry members of our trial site, perovskite and bcc alloy, yield high degeneracy figures (5.9\% and 4.9\%, respectively). However, symmetry and degeneracy do not appear to correlate, as sphalerite degeneracy is very low (0.29\%) despite its high symmetry, and gehlenite has a high degeneracy rate (2.9\%) despite the asymmetric origin unit cell. In any case, the occurrence of degenerate virtual cells can be solved by simply eliminating said cells from the sample set, at little cost to the accuracy as determined by the error margins.

\begin{figure}[htp]
\centering
\includegraphics[width=0.6\textwidth]{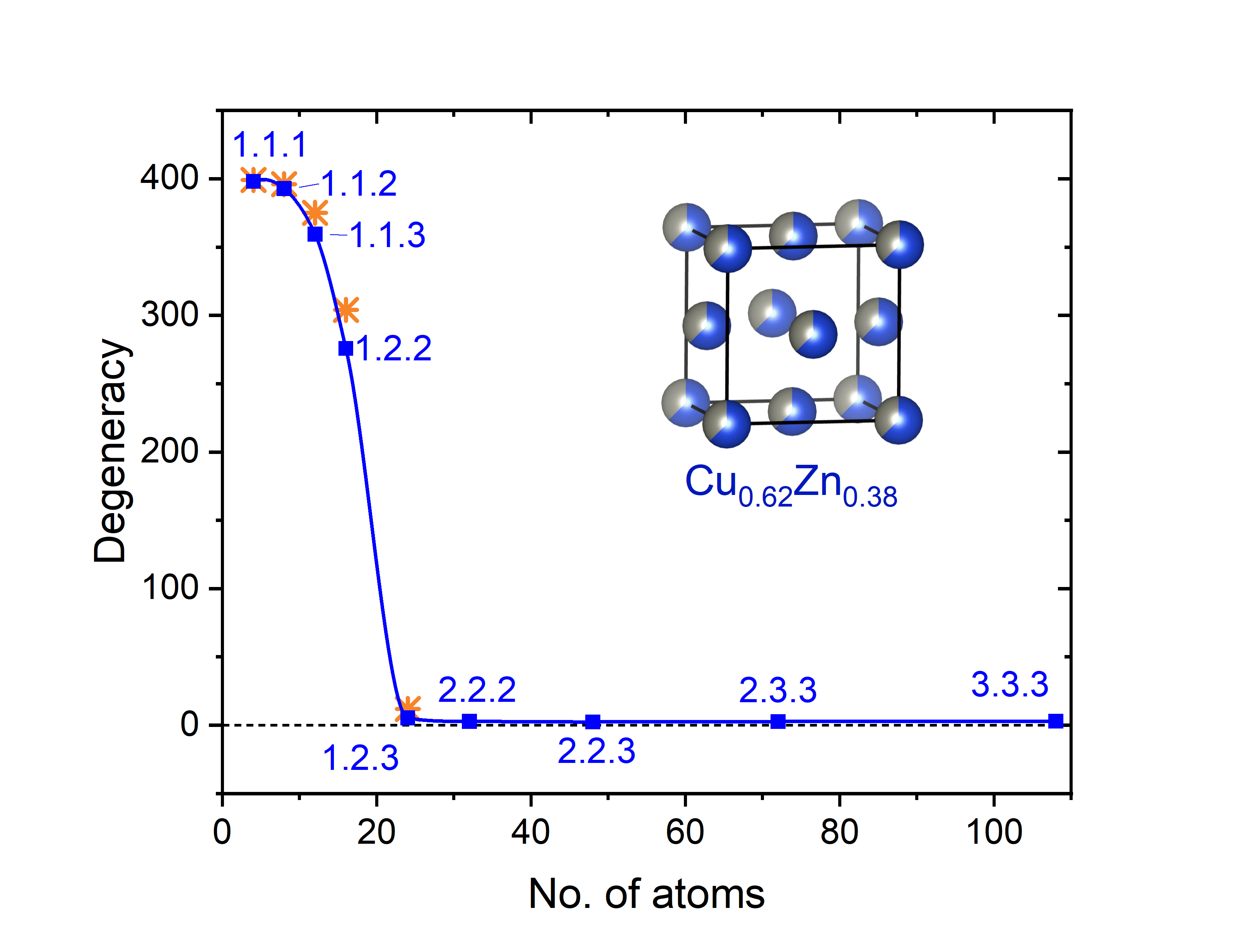}
\caption{In blue: degeneracy (in number of cells out of 400) of virtual cell sample sets of conventional brass, Cu$_{0.62}$Zn$_{0.38}$\cite{owen_x-ray_1923}. These are taken as the mean of 30 runs for each data point; error bars are too small to be discernible in this graph. In orange asterisks: number of redundant structures according to the symmetry equivalence checker implemented in the Atomic Simulation Environment package (ASE)\cite{lonie_identifying_2012, HjorthLarsen2017}.}
\label{fig:degeneracy}
\end{figure}

It would be interesting to examine the choice of the supercell to the degeneracy rate of the virtual cell sample set. For this, we chose to use as an example the high-symmetry cell of brass (Cu$_{0.62}$Zn$_{0.38}$)\cite{owen_x-ray_1923}. Although it would seem that choosing an asymmetric unit cell (e.g. $1 \times 2 \times 3$) would tend to produce less degeneracy than a more isotropic multiplicity (e.g. $2 \times 2 \times 2$), we find that supercell size is a much stronger predictor of virtual cell degeneracy, and any brass supercell with more than 20 atoms would give a degeneracy of $\sim 1\%$ or less (Figure \ref{fig:degeneracy}). A relatively small $2 \times 2 \times 2$ supercell is sufficient to generate 400 virtual cells with very little redundancy rate, about 0.7 \%, which would justify using the sample set as is. For comparison, the ``rule of thumb'' of a minimum of 15$\si{\angstrom}$ between periodic boundary images would require a supercell size of $5 \times 5 \times 5$.

The strategy of testing virtual cell redundancy through CHGNet total energy is tested against the symmetry equivalence checker, implemented in the Atomic Simulation Environment (ASE) package \cite{lonie_identifying_2012, HjorthLarsen2017}. Owing to the obligatory threshold setting in the ASE symmetry equivalence checker, it is more permissive and thus tends to identify more symmetrically equivalent pairs than the CHGNet total energy criterion. Among the $1 \times 2 \times 2$ virtual cells, the ASE symmetry equivalence checker identifies 949 equivalent pairings, of which 782 are also identified using CHGNet total energy comparisons (there are no pairings identified by CHGNet total energy comparisons alone). For $2 \times 2 \times 3$ supercells or larger, the calculation time of the ASE symmetry equivalence checker becomes prohibitive, highlighting the necessity of the CHGNet total energy method in high-throughput checking for symmetric redundancies within a set of large virtual cells.

\subsection*{Calculation times}

\begin{figure}[htp]
\centering
\includegraphics[width=0.6\textwidth]{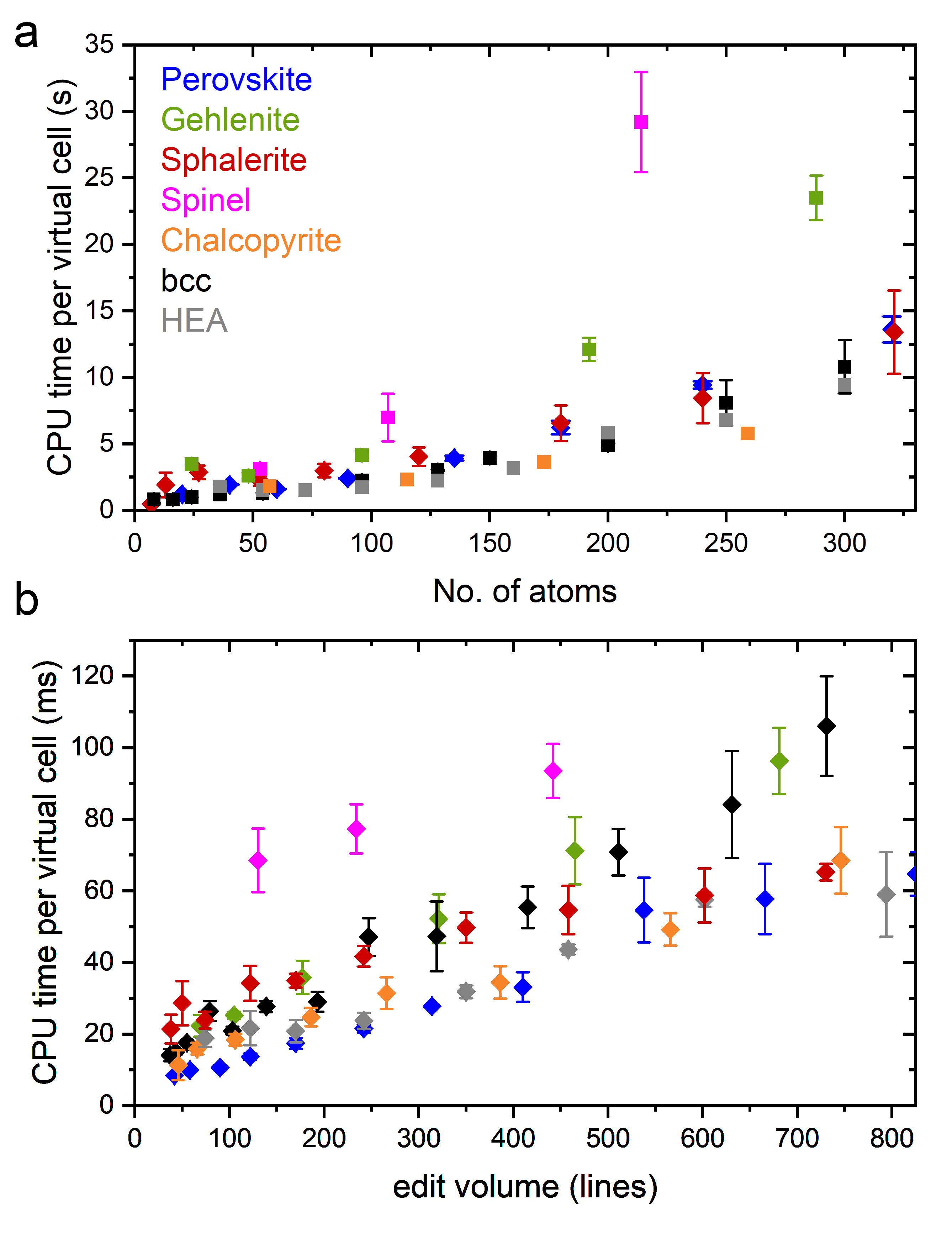}
\caption{(a) CPU time used to generate and structurally optimize a virtual cell against number of atoms; (b) CPU time used to generate a virtual cell only, against edit volume.}
\label{fig:calctimes}
\end{figure}

The generation of virtual cells with structural optimization on the structures in the trial set consumed CPU time of a magnitude of $\sim$ 1-10 s (Figure \ref{fig:calctimes} a). The calculation load correlates linearly with the number of atoms in the virtual cell for cells of $\sim$300 atoms, suggesting that structural optimization is the predominant contributor to the calculation load. 

When structural optimization is discounted, the CPU time requirement drops to $\sim$10-100ms (Figure \ref{fig:calctimes}b). By necessity given the sorting process, the calculation time is correlated with the edit volume (in units of lines), defined as (edit block size) $\times$ (supercell multiplicity) $+$ (number of lines in the file), where the edit block size represents the number of lines in the supercell file corresponding to an instance of a disordered site.

\subsection*{Comparison with first principles calculations}

We evaluate the potential trade-offs in accuracy from substituting DFT with CHGNet. In addition, since electronic structural attributes such as density of states (DOS) are currently inaccessible from MLIPs alone, we also take the opportunity to examine the variations of DOS possible by the randomized assignment of atomic species to disordered sites.

\begin{figure}[htp]
\centering
\includegraphics[width=\textwidth]{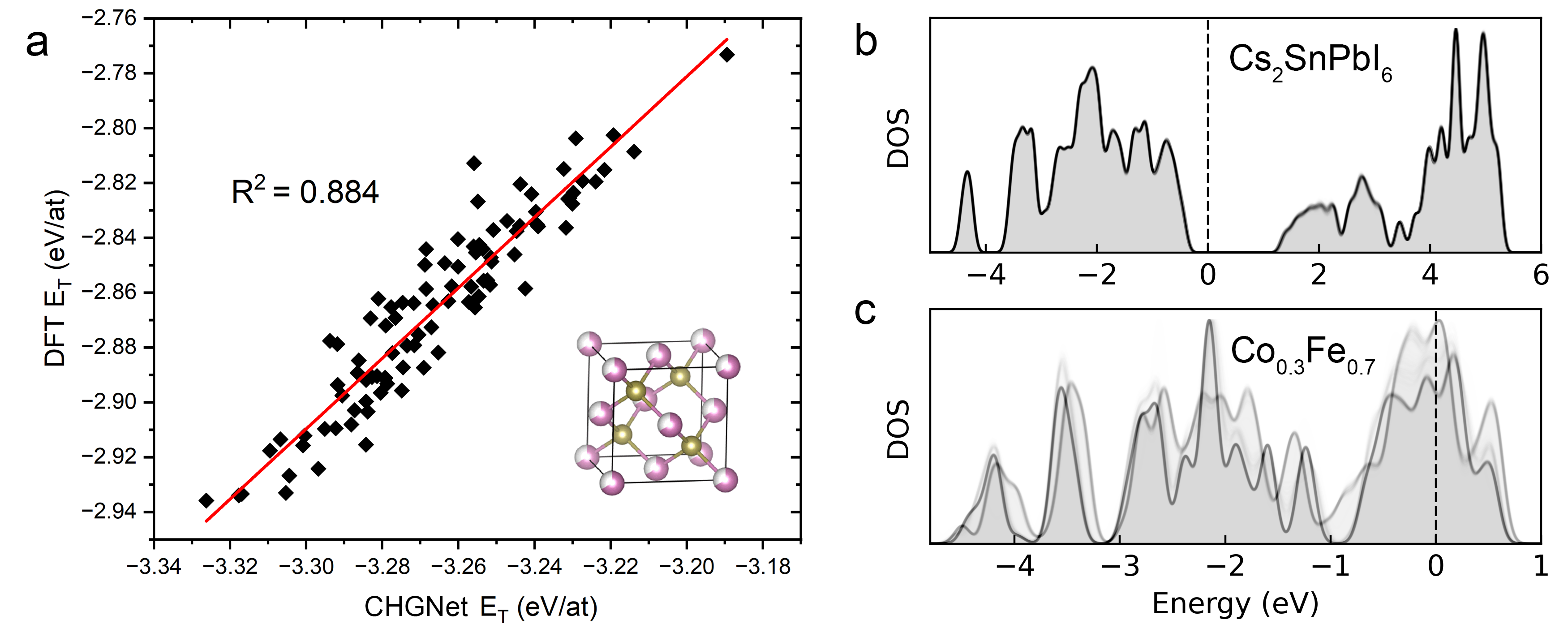}
\caption{(a) Correlation between CHGNet- and DFT-derived total energy of 100 un-relaxed sphalerite (In$_2$Te$_3$) virtual cells. (b-c) Overlay density of states (DOS) graphs of 400 virtual supercells for bcc alloy (Co$_{0.3}$Fe$_{0.7}$) and perovskite (Cs$_2$SnPbI$_6$). DOS graphs are plotted using the \texttt{sumo} package\cite{m_ganose_sumo_2018}.}
\label{fig:vaspdos}
\end{figure}

First, we test the accuracy of CHGNet total energy predictions on a set of 100 un-relaxed sphalerite (In$_2$Te$_3$) virtual cells against DFT-calculated total energies. In this case, we find that the CHGNet total energy correlates strongly with the DFT total energy, with $R^2 = 0.884$ (Figure \ref{fig:vaspdos}a).

\begin{table}[h]
\begin{center}
\caption{Boltzmann-averaged density $\rho$ and electronic band gap $E_g$ of bcc alloy (Co$_{0.3}$Fe$_{0.7}$) and perovskite (Cs$_2$SnPbI$_6$), as calculated with \texttt{virp} (powered with CHGNet for $\rho$ and \texttt{matgl} for $E_g$), compared to first principles calculations with VASP.}\label{tbl:vsdft}%
\begin{tabular}{@{}llllll@{}}
\toprule
System & Property & ($E_g$ model) & CHGNet & VASP & Error \\
\midrule
bcc alloy & $\rho$ (g/cm$^3$) & & 8.11 & 8.12 & $-0.01$ \\
\multirow{5}{*}{Perovskite} & $\rho$ (g/cm$^3$) & & 4.43 & 4.49 & $-0.06$ \\
& \multirow{4}{*}{$E_g$ (eV)} & PBE & 0.92 & \multirow{5}{*}{1.06} & $-0.14$ \\
& & GLLB-SC & 2.62 & & $+1.56$ \\
& & HSE & 1.24 & & $+0.18$ \\
& & SCAN & 1.22 & & $+0.16$ \\
\botrule
\end{tabular}
\end{center}
\end{table}

Next, we obtained CHGNet- and DFT-derived material properties for bcc alloy (Co$_{0.3}$Fe$_{0.7}$) and perovskite (Cs$_2$SnPbI$_6$), summarized in Table \ref{tbl:vsdft}. Boltzmann averaging and CHGNet structural optimization produced under-predicted densities of 0.01 g/cm$^3$ (bcc alloy) and 0.06 g/cm$^3$ (perovskite) as compared to DFT results. Electronic band gap predictions with \texttt{matgl} are more variable, with the PBE model producing the lowest error (-0.14 eV) out of the 4 models compared to DFT. The variability of \texttt{matgl} reflects the fact that foundation models of electronic band gap prediction are still in the early stages of development, and may need further improvements before they can be used reliably in high-throughput evaluations. The low error in the densities can be thought of as a result of Boltzmann averaging and the small spread of total energies.

We examine the effect of disorder on electronic DOS by overlaying the DOS graphs to obtain a picture of an ``average'' DOS (Figure \ref{fig:vaspdos}b-c). The opacity of the individual supercell DOS can be weighted according to the relative formation energies of the supercells, although if the individual supercell DOS are equally weighed, the difference this makes in the average DOS is negligible. This reinforces our observation that the spread in formation energies is small among the virtual cells. The overlap in supercell DOS in perovskite is almost perfect. In contrast, bcc alloy exhibits a variability in DOS profile among its supercells. Two explanations are possible here: first, it is possible that this is another instance of a spurious phase bifurcation originating from periodic boundary effects, as we have seen before in AgSbTe$_2$. Secondly, this may point to a tendency in metallic alloys to exhibit short range order\cite{sheriff_quantifying_2024}, implying that the random site filling paradigm of \texttt{virp} may not reflect the preferred state of the material with a limited degree of order. This explanation would relegate bcc alloy Co$_{0.3}$Fe$_{0.7}$ to the category of correlated disorder materials, which will be the focus of our following studies.

\section*{Discussion}\label{sec12}

This work builds on recent advances in the modeling of site-disordered crystal structures and addresses some of its most pertinent problems. The coherent potential approximation (CPA) is limited in its applicability to just metal alloys. Cluster expansion-based methods are best implemented for simpler site-disorder systems, and often do not cater for multiple disordered sites. Thus, random permutative filling and statistical post-processing have the advantage of being more generally applicable than the other methods, as illustrated in our trial set.

Earlier authors\cite{oses_aflow_2023} have asserted that a exhaustive sampling of virtual structures for a given supercell is necessary for accurate predictions of material properties. To address this, we have shown that this assertion is not necessarily true: According to Yamane sampling theory, for a given error margin, a sample size of a few hundred models can represent the properties of the complete set of virtual supercells, even when the size of configurational space becomes very large (we have seen as high as $10^{110}$ in our trial set). An error margin of 5\% would prescribe for us a sample size of $\sim$400, although multimodalities in the material properties of the sample set may produce a higher error margin than aimed.

Yamane theory, MLIP structural optimization, and machine learning models of property prediction have the potential to accelerate greatly the prediction of the structure and properties of virtual cells (and, by extension, the site-disordered crystal) compared to first principles simulations. While a DFT calculation can last from hours to days, CHGNet calculations can last on the order of tens of seconds, and the random site filling operation can be done on the order of tens of milliseconds per virtual cell. Though may be systematic errors inherent from the underlying neural network model of the MLIP, future improvements on these models can be integrated into \texttt{virp} as well. 

Existing methods may also be hampered by computationally intensive symmetry analysis of virtual cells aimed at identifying redundant structures, especially in larger supercells\cite{okhotnikov_supercell_2016}. Here, we propose the simpler solution of comparing CHGNet total energies \emph{post hoc} to find and eliminate equivalent structures. In the example of the brass crystal structure, we demonstrate that larger supercells would bring down the virtual cell degeneracy rates to about $\sim 1\%$. Larger supercells are crucial to minimizing errors induced by the periodic boundary. In contrast to \texttt{Supercell}, \texttt{virp} incentivizes the user to define larger supercells, not smaller ones, which is important to eliminate spurious artifacts arising from interactions across the periodic boundary.

Random permutative site filling is not powerful enough to create virtual cells with correlated disorder, as the site filling in each disordered site could have a dependency in the filling (or lack thereof) of a neighboring site. Thus, many common materials, such as water ice, cannot be adequately treated, as the sample set would be populated by invalid virtual cells. This dependency must be inferred, as it is not encoded in the CIF file. The use of new methods to generate large numbers of valid virtual cells for a correlated disorder crystal structure will be a topic for a future study.

\section*{Methods}\label{sec11}

\subsection*{Boltzmann averaging}

Similar to \texttt{aflow++}, the Boltzmann-averaged expectation value $\left < P \right >$ of a certain property $P$ from the predicted property $p_i$ of each virtual cell (of energy $E_i$) in the sample set can be determined through the equation
\begin{equation}
    \left < P \right > = \frac{\sum_i p_i e^{\frac{-E_i}{k_{B} T}}}{\sum_j e^{\frac{-E_j}{k_{B} T}}}
\end{equation}
where $k_{B}$ is the Boltzmann constant.

\subsection*{Virtual cell generation}

Database building operations are performed on the trial sets. The trial crystal structures are read using the text-based crystallographic information file (CIF) format. For each trial cell, sets of 400 ($n_Y$ for $N \rightarrow \infty$ and $e=5\%$) or 1300 (to evaluate $n_Y=286$) virtual cells are generated, structurally optimized using \texttt{CHGNet}\cite{deng_chgnet_2023}, and assigned a predicted band gap using \texttt{matgl}\cite{chen_learning_2021}. Each batch operation can be completed in the space of a few weeks.

When \texttt{virp} treats a site-disordered unit cell (Figure \ref{fig:virpintro}a), it first creates a supercell by replicating the unit cell a number of times along the three crystallographic axes; the number chosen should be large enough to minimize periodic boundary effects. From the supercell, we generate a virtual cell (Figure \ref{fig:virpintro}b) with a stoichiometry which closely matches the original.

Following the method of \texttt{Supercell}, each instance in each disordered site in the supercell is randomly assigned an atom (or vacancy) according to the proportional occupancy of the elements in the site. This is done by establishing a site assignment procedure: First, we map the occupancies onto a discrete array (a ``snap'') by rounding the cumulative occupancies to the nearest integer (Figure \ref{fig:virpintro}c). Secondly, we guarantee that each element has at least one assigned site. Lastly, if the cumulative occupancy is exactly in the middle of two integers, the antibiasing mechanism rounds it up or down in the snap with equal probability, as long as the second condition can be fulfilled (Figure \ref{fig:virpintro}d). The total number of distinct virtual cells $N_v$ is thus:

\begin{equation}
N_{v}=\prod_{s}\sum_{\mathbf{x}}\frac{m(s)!}{\prod_{i} x_{i} !}.
\end{equation}

Here, $s$ denotes a crystallographic site, $m(s)$ is the multiplicity of the site in the supercell, and $\mathbf{x} = (x_1, x_2, \dots, x_n) \in \mathbb{N}^n$ denotes a snap.

\subsection*{Prediction of properties}

For each material, we predict two material properties. First, the density is obtained from the structurally optimized virtual cell. Second, band gap predictions are made in the \texttt{matgl} package based on the machine-learned \texttt{MEGNet} band gap models \cite{chen_learning_2021}, which are based on the PBE, GLLB-SC, HSE, and SCAN functionals. It is important to note that CHGNet structural optimization is limited in accuracy relative to DFT. These inaccuracies are compounded in the \texttt{MEGNet} band gap prediction, which is created to estimate band gaps in high-throughput screening rather than as an accurate assessment of band gap in individual materials. Thus, we should expect a degree of systematic error in our predictions.

\subsection*{Measuring calculation times}

The CPU time expended in the generation of one cell are calculated using the \texttt{timeit} software. An average expended CPU time is obtained from 7 code block calls. The calculations are performed on a server with 2 Intel Xeon Gold 6336Y CPU processors, with 24 cores each, 256GB DDR4 RAM, and a NVIDIA A40 GPU with 10752 CUDA cores.

\subsection*{First principles calculations}

First principles calculations were performed using density functional theory (DFT) as implemented in the Vienna Ab initio Simulation Package (VASP) \cite{Kresse1993,Kresse1994,Kresse1996,Kresse1996a}. The projector augmented-wave (PAW) method\cite{Blochl1994,Kresse1999} was used to describe the core–valence interaction, with a plane-wave energy cutoff of 520eV. Exchange-correlation effects were treated within the generalized gradient approximation (GGA) of Perdew, Burke, and Ernzerhof (PBE)\cite{Perdew1996}.

Structural optimization was performed using a two-step relaxation scheme and the conjugate gradient algorithm for ionic updates. In the first step, a low-precision relaxation was performed with an electronic convergence threshold of 0.1 meV to accelerate convergence toward a reasonable structure. In the second step, full high-precision relaxation was performed with an electronic convergence threshold of 1 $\mu$eV. The ionic convergence threshold, in terms of force on atoms, is maintained at 0.02 eV/$\si{\angstrom}$. All structural parameters, including atomic positions, lattice shape, and volume, are relaxed. 

A two-step approach was employed for accurate DOS analysis. First, a scalar-relativistic self-consistent field (SCF) calculation was performed to generate a converged charge density The resulting charge densities were then used in a second static calculation to obtain the total and projected DOS. In these steps, the electronic convergence threshold is also set at 1 $\mu$eV. In the case where atoms heavier than Sn are involved, we also include spin–orbit coupling (SOC) to account for relativistic effects during the second step.

In all above calculations, reciprocal sampling was done using a single $\Gamma$ point, taking into account the largeness of the supercells, and a Gaussian smearing of width 0.1 eV. These settings are designed to be broadly applicable across many types of materials, whether conducting and insulating.

\backmatter



\bmhead{Acknowledgments}

We acknowledge Savyasanchi Aggarwal for his contributions in our fruitful discussions. A. P. C. and K. H. acknowledge support from Ministry of Education (MOE) Academic Research Fund (AcRF) Tier 1, Sponsor Award ID RG138/23. Calculations are performed on the \texttt{Khompute} server in the School of Materials Science and Engineering, Nanyang Technological University, with assistance from Nong Wei, and on the ASPIRE 2A supercomputer at the National Supercomputing Centre (NSCC), Singapore, with assistance from Dr. Nicholas Cheng Lin Quan.

\section*{Declarations}

\bmhead{Competing interests}

There are no competing interests to declare.

\bmhead{Data availability}

Data generated and analyzed in this study are available on Zenodo (\url{https://zenodo.org/records/16666679}) and GitHub (\url{https://github.com/Kedar-Materials-by-Design-Lab/virp-data}; excluding VASP calculation folders) repositories. Crystallographic data downloaded from ICSD are excluded from the available data in compliance with ICSD data policy.

\bmhead{Code availability}

The underlying code for this study (\texttt{virp}) is available in GitHub (\url{https://github.com/andypaulchen/virp}) and PyPI (\url{https://pypi.org/project/virp/}). Virtual cells in this study are generated with version \texttt{v1.2.1}.

\bmhead{Authors' contributions}

\textbf{A. P. C.} conceptualized the project, created and tested the software, performed first principles calculations, and wrote the original draft of the manuscript. \textbf{M. H. P.} co-created and tested the software, and also edited the manuscript. \textbf{K. H.} conceptualized and supervised the project, acquired funding, and edited the manuscript.









\bibliography{references}


\end{document}